\documentclass{article}
\usepackage{spconf,amsmath,graphicx}
\usepackage{hyperref}
\usepackage[utf8]{inputenc}
\usepackage{amssymb}
\usepackage[ruled,vlined]{algorithm2e}  
\usepackage[T1]{fontenc}
\usepackage{xcolor}
\usepackage{float}

\let\OLDthebibliography\thebibliography
\renewcommand\thebibliography[1]{
  \OLDthebibliography{#1}
  \setlength{\parskip}{1.55pt plus 0.5pt minus 0.5pt}
  \setlength{\itemsep}{1.55pt plus 0.5pt minus 0.5pt}
}

	\usepackage{pgfplots}
  \pgfplotsset{compat=newest}
\newlength\figurewidth
\usetikzlibrary{patterns}

\makeatletter
\newcommand{\removelatexerror}{\let\@latex@error\@gobble}
\makeatother

\def\RR{\mathbb{R}}

\def\CC{\mathbb{C}}

\DeclareTextAccent{\ring}{OT1}{23}

\def\x{\vect{x}}
\def\u{\vect{u}}

\def\z{\vect{z}}

\def\p{\vect{p}}
\def\q{\vect{q}}
\def\c{\vect{c}}

\newcommand{\Id}{\mathrm{Id}}
\newcommand{\syn}{\adjoint{\ana}}
\newcommand{\ana}{A}

\newcommand{\xq}{\vect{x}^\mathrm{q}}
\newcommand{\ASet}{\Gamma}
\newcommand{\SSet}{\Gamma^*}

\newcommand{\Tdim}{P}
\newcommand{\TFdim}{Q}
\newcommand{\const}{\alpha}

\newcommand{\norm}[1]{\|#1\|}

\newcommand{\adjoint}[1]{#1^*}

\newcommand{\vect}[1]{\mathbf{#1}} 

\newcommand{\argmin}{\mathop{\operatorname{arg~min}}}

\newcommand{\prox}{\mathrm{prox}}
\renewcommand{\approx}{\mathrm{approx}}
\newcommand{\soft}{\mathrm{soft}}
\newcommand{\proj}{\mathrm{proj}}

\newcommand{\sgn}{\mathrm{sgn}}

\newcommand{\dsdr}{\ensuremath{\Delta\text{SDR}}}

\title{Audio dequantization using (co)sparse (non)convex methods}
\name{Pavel Z\'avi\v{s}ka, Pavel Rajmic, Ond\v{r}ej Mokr\'y\sthanks{The work was supported by the project 20-29009S of the Czech Science
Foundation (GA\v{C}R).}}
\address{Signal Processing Laboratory, Brno University of Technology, Brno, Czech Republic}

\begin{document}
\maketitle
\begin{abstract}
The paper deals with the hitherto neglected topic of audio dequantization.
It reviews the state-of-the-art sparsity-based approaches and proposes several new methods.
Convex as well as non-convex approaches are included, and 
all the presented formulations come in both the synthesis and analysis variants. 
In the experiments the methods are evaluated
using the signal-to-distortion ratio (SDR) and PEMO-Q, a~perceptually motivated metric.
\end{abstract}

\section{Introduction}
\label{sec:intro}

\vspace{-1ex}
The task of dequantization is to recover a~signal from its quantized observation.
Quantization is a~nonlinear distortion that introduces perceptually unpleasant artifacts.
Practically,
a~digital signal is always quantized, but normally the bit depth of each audio sample is so high
(at least 16 bits per sample)
that the effect of quantization is imperceptible.
However, once the bit allocation starts to decrease, the quantization starts to be pronounced.
Using a~low bit depth in audio may be forced
(the bandwidth limitation in communication systems
\cite{BrauerGerkmannLorenz2016:Sparse.reconstruction.of.quantized.speech.signals}, an inadequate setup during music recording)
or intentional
(compression of the signal via
straightforward
bit depth reduction, mild requirements on the quality).
The time-domain quantization actually fits into a~more complex audio coding task;
recently, a~novel audio coding strategy has been proposed with quantization allowed in multiple domains
\cite{MokryRajmicZaviska2020:Flexible.framework.audio.reconstruction}.

In the past decade, audio processing methods exploiting the sparsity of audio signals
has drawn much attention.
This is also true for the area of audio dequantization,
but it is clear from a~look into the literature that the interest in this task is much weaker than the attention paid to the closely related reconstruction problems of
audio declipping
\cite{ZaviskaRajmicOzerovRencker2021:Declipping.Survey,SiedenburgKowalskiDoerfler2014:Audio.declip.social.sparsity,Kitic2015:Sparsity.cosparsity.declipping}
and audio inpainting (i.e., filling the missing gaps in the signal)
\cite{Adler2012:Audio.inpainting,MokryRajmic2020:Inpainting.revisited,Marafioti2019:DNN.inpainting}.

The quantization limits the number of possible values the signal can attain;
each original signal sample is rounded to the nearest quantization level
\cite{Zolzer2011:DAFX}.
%
In this study, we stick to the uniform quantizer for simplicity,
although non-uniform quantizers
\cite{ITU-T_G.711} could be considered without loss of generality.
In the case of the standard, so-called mid-riser uniform quantizer,
all the quantization levels are equally distributed. 
The quantization step is $\Delta = 2^{-w+1}$, with $w$ denoting the word length in bits per sample (bps). 
Specifically, the quantized signal $\xq \in \RR^N$ is obtained according to the formula
\begin{equation}
(\xq)_n = \sgn^+ (\x_n)\cdot \Delta \cdot \left( \left\lfloor \frac{|\x_n|}{\Delta} \right\rfloor + \frac{1}{2} \right),
\label{eq:uniform_quantization}
\end{equation}
where the $n$-th sample of the signal is indicated by the index $n$,
and $\sgn^+(z)$ returns $1$ for $z\geq0$ and $-1$ for $z<0$.


For the sake of
reviewing the state of the art,
it will be convenient to discuss the issue referred to as the \emph{solution consistency}.
The solution to a~dequantization problem is called consistent with the observed samples
if each sample of the dequantized signal lies within the same quantization interval as the respective original sample does.
If this is not true, the solution is called inconsistent.

Brauer et al.\
\cite{BrauerGerkmannLorenz2016:Sparse.reconstruction.of.quantized.speech.signals}
approximate the dequantization by formulating a~convex optimization problem, solved by the proximal splitting algorithms
\cite{combettes2011proximal,ChambollePock2011:First-Order.Primal-Dual.Algorithm}.
They work exclusively on speech signals and no computer implementation is available, unfortunately.
The method of 
\cite{BrauerGerkmannLorenz2016:Sparse.reconstruction.of.quantized.speech.signals}
is consistent with the quantized observations.

Záviška et al.\ 
\cite{ZaviskaRajmic2020:Dequantization}
follow up on the just mentioned study.
They continue in exploring the potential of sparsity-based methods in reconstructing quantized audio.
The authors reimplement the method of
\cite{BrauerGerkmannLorenz2016:Sparse.reconstruction.of.quantized.speech.signals}
but in particular, they significantly extend the range of the evaluation scenarios
and introduce also the analysis (cosparse) model.
They also show
that using the Gabor transform in place of the discrete cosine transform used in
\cite{BrauerGerkmannLorenz2016:Sparse.reconstruction.of.quantized.speech.signals}
leads to improved results.

Rencker et al.\
\cite{RenckerBachWangPlumbley2018:Fast.iterative.shrinkage.declip.dequant-iTwist18,RenckerBachWangPlumbley2018:Sparse.recovery.dictionary.learning}
allow inconsistency of the solution.
However, the inconsistency is penalized, leading to an unavoidable existence of a~user-defined parameter
that balances the degrees of sparsity and inconsistency of the solution.
Their formulation also leads to a~convex problem,
but now admitting a~quick solver,
which comes from the fact that the inconsistency penalty is a~smooth function.

Even more recently, a~deep-learning-based method has been published \cite{BrauerZhaoLorenzFingscheidt2019:Dequantization_speech_signals}.
Nonetheless, the classical approaches have not yet been fully explored,
which is why the goal of this paper is to evaluate
a~number
of sparsity-based audio dequantization methods.
In the paper, we cover the methods mentioned above but also propose several brand new approaches.

\section{Sparsity-based formulation}
\label{sec:sparsity_ftw}
The above introduced consistent solution is formally characterized such that it belongs to the set 
$\ASet\subset\RR^\Tdim$,
where 
\begin{equation}
	\ASet = \{\x\in\RR^\Tdim \mid \norm{{\x - \xq}}_\infty < \Delta/2\}.
	\label{eq:gamma_analysis}
\end{equation}
In words, the dequantized sample cannot lie outside of the quantization interval that corresponds to the original sample.
The set $\ASet$ is a~convex, multidimensional interval, or a box,
which indicates the
relationship to audio inpainting and declipping \cite{MokryRajmicZaviska2020:Flexible.framework.audio.reconstruction}.
This observation also motivates the adaptation of efficient
reconstruction algorithms to dequantization.

Given $\xq$, there are numerous feasible signals in $\ASet$, making the reconstruction ill-posed.
Sparsity can come into play here
as a~regularizer.
In audio, the assumption of sparsity of the time-frequency coefficients of the signal is usually employed.
Denote $\c\in\CC^\TFdim$ a vector of coefficients and $\x\in\RR^\Tdim$ a time-domain signal.
The regularized problem
%
%
is then to find $\x\in\ASet$ such that the corresponding coefficients $\c$ are as sparse as possible.
The time-frequency transforms are assumed redundant
(therefore $Q>P$),
leading to two distinct variants based on the relationship between $\x$ and $\c$
\cite{NamDaviesEladGribon2013:CosparseAnalysisModel}:
%
\begin{subequations}
	\label{eq:sparsity.formulation:anasyn}
	\begin{align}
	&\argmin_{\c\in\CC^\TFdim} \norm{\c}_0\quad & \text{subject to}\quad & \syn\c\in\ASet,
	\label{eq:sparsity.formulation:syn}\\ 
	&\argmin_{\x\in\RR^\Tdim} \norm{\ana\x}_0\quad & \text{subject to}\quad & \x\in\ASet.
	\label{eq:sparsity.formulation:ana}
	\end{align}
\end{subequations}
%
%
The symbol $\norm{\cdot}_0$ returns the number of non-zero elements in the vector, i.e., sparsity.
Since the formulation \eqref{eq:sparsity.formulation:syn} uses the operator $\syn$ called synthesis,
it is referred to as the \emph{synthesis formulation}.
Similarly, \eqref{eq:sparsity.formulation:ana}
is the \emph{analysis formulation}.

The condition $\syn\c\in\ASet$ in Eq.\,\eqref{eq:sparsity.formulation:syn}
may be equivalently written as $\c\in\SSet$ with
$\SSet = \{ \c'\in\CC^\TFdim \mid \syn\c' \in \ASet \}$.
The crucial observations are that both the sets $\ASet$ and $\SSet$ are convex and that there exist explicit formulas for the corresponding projection operators \cite{RajmicZaviskaVeselyMokry2019:Axioms}.
This fact is utilized by the numerical solvers described in Sec.\,\ref{sec:algorithms}.

Throughout the rest of the paper, we assume the use of Parseval tight frames,
i.e.\ linear operators for which it holds
$\syn\ana=\Id$ \cite{christensen2008}.
Any tight frame can be simply scaled to become Parseval-tight.

\section{Problems and Algorithms}
\label{sec:algorithms}

The presence of the $\ell_0$ penalty makes 
the problems in \eqref{eq:sparsity.formulation:anasyn} NP-hard
and therefore the solutions need to be approximated in practice.
This section is devoted to the presentation of formulations that follow from different means of approximation,
and the corresponding algorithms.

\subsection{Consistent \texorpdfstring{$\ell_1$}{l1} minimization}
\label{ssec:synthesis.consistent.l1}
The first option is to use the
$\ell_1$ norm instead of $\ell_0$ to make the whole problem convex
\cite{DonohoElad2003:Optimally,FornasierEditor2010:SparseRecoveryBook},
allowing the use of convex optimization
\cite{chen,Boyd:2004:Book:ConvexOptimization}.

The convex relaxation of \eqref{eq:sparsity.formulation:syn} reads
\begin{equation}
	\argmin_{\c\in\CC^\TFdim} \norm{\c}_1\quad \text{subject to}\quad \syn\c\in\ASet.
	\label{eq:synthesis.l1.const}
\end{equation}
To solve \eqref{eq:synthesis.l1.const}
numerically, proximal splitting methods \cite{combettes2011proximal} offer a~suitable choice.
%
We show in
\cite{ZaviskaRajmic2020:Dequantization}
that the solution to
\eqref{eq:synthesis.l1.const}
can be found via the Douglas--Rachford (DR) algorithm \cite[Sec.\,4]{combettes2011proximal}.
The DR algorithm is summarized in Alg.\,\ref{alg:DR_dequantization}.
Bearing in mind that $\syn\c\in\ASet$ is equivalent to $\c\in\SSet$,
the DR algorithm is derived such that its first step uses the projection onto the set $\SSet$.
The projection can be computed as \cite{RajmicZaviskaVeselyMokry2019:Axioms}
\begin{equation}
	\proj_{\SSet}(\c) =  \c + \ana\left(\proj_{\ASet}(\syn\c)-\syn\c\right).
	\label{eq:proj.on.sset}
\end{equation}
The second step of the DR algorithm uses $\soft_{\const}$, the soft thresholding operator with a~threshold $\const > 0$
\cite{DonohoElad2003:Optimally,Beck2017:First.Order.Methods}.

\begin{figure}[t]
\begin{minipage}[t]{\columnwidth}
\removelatexerror
\begin{algorithm}[H]
	\DontPrintSemicolon
	\SetAlgoVlined
	\small
	\KwIn{Set starting point $\z^{(0)} \in \CC^\TFdim$ and
				parameter $\gamma > 0$.
	}
	\For{$i=0,1,\dots$\,}{
		$\c^{(i)} = \proj_{\SSet}(\z^{(i)})$ \\
		$\z^{(i+1)} = \z^{(i)} + \soft_{\gamma}(2\c^{(i)} - \z^{(i)}) - \c^{(i)}$ \\
	}
	\KwRet{$\c^{(i)}$}
	\caption{\mbox{Douglas--Rachford algorithm solving \eqref{eq:synthesis.l1.const}}}
	\label{alg:DR_dequantization}
\end{algorithm}

\begin{algorithm}[H]
	\DontPrintSemicolon
	\SetAlgoVlined
	\small
	\KwIn{Set starting points $\p^{(0)} \in \RR^\Tdim, \q^{(0)} \in \CC^\TFdim$. \\
				Set parameters $\zeta, \sigma > 0$, $\zeta\sigma\norm{\ana}^2<1$, and $\rho\in[0,1]$.
	}
	\For{$i=0,1,\dots$\,}{ 
		$\q^{(i+1)} = \mathrm{clip}_{1}(\q^{(i)}+\sigma \ana \x^{(i)})$\\
		$\p^{(i+1)} = \proj_{\ASet}(\p^{(i)}-\zeta \syn{\q}^{(i+1)})$\\
		$\x^{(i+1)} = \p^{(i+1)} + \rho (\p^{(i+1)} - \p^{(i)}) $\\
	} 
	\KwRet{\(\p^{(i+1)}\)}
	\caption{\mbox{Chambolle--Pock algorithm solving \eqref{eq:analysis.l1.const}}}
	\label{alg:CP_dequantization}	
\end{algorithm}
\end{minipage}
\vspace{-1.5em}
\end{figure}

Similar to the synthesis model, we relax
\eqref{eq:sparsity.formulation:ana} as
\begin{equation}
	\argmin_{\x\in\RR^\Tdim} \norm{\ana\x}_1\quad \text{subject to}\quad \x\in\ASet.
	\label{eq:analysis.l1.const}
\end{equation}
Compared with \eqref{eq:synthesis.l1.const}, the analysis problem \eqref{eq:analysis.l1.const} is more challenging because of the presence of the composite function $\norm{\ana\,\cdot}_1$.
The proximal operator of such a composition is not available even in the case of a~Parseval tight frame in place of $\ana$ \cite{MokryRajmic2020:Approximal.operator}.
However, the Chambolle--Pock (CP) algorithm \cite{ChambollePock2011:First-Order.Primal-Dual.Algorithm}
can handle such an optimization problem.
%
The CP algorithm for dequantization is summarized in Alg.\,\ref{alg:CP_dequantization}.
In the algorithm, $\mathrm{clip}_{\const} = \Id-\soft_{\const}$.

\subsection{SPADQ, non-convex minimization}
\label{ssec:spadeq}
Another option in approximating the problems \eqref{eq:sparsity.formulation:anasyn}
is to relax the strict relationship of $\x$ and $\c$ and approach the problem heuristically
\cite{Kitic2015:Sparsity.cosparsity.declipping,ZaviskaRajmicMokryPrusa2019:SSPADE_ICASSP}.
The acronym SPADQ stands for the Sparse Audio Dequantizer, a~natural adaptation of the Sparse Audio Declipper (SPADE)
\cite{Kitic2015:Sparsity.cosparsity.declipping, ZaviskaRajmicMokryPrusa2019:SSPADE_ICASSP}
or Inpainter (SPAIN)
\cite{MokryZaviskaRajmicVesely2019:SPAIN}
to the task of audio dequantization.
The dequantization models are derived from a~reformulation of \eqref{eq:sparsity.formulation:anasyn}
with less rigid coupling of the signal and the coefficients, governed by the parameter $\epsilon$.
Based on \cite{Kitic2015:Sparsity.cosparsity.declipping, ZaviskaRajmicMokryPrusa2019:SSPADE_ICASSP},
three options are available:
\begin{subequations}
	\label{eq:spadeq.formulation:anasyn}
	\begin{align}
		\argmin_{\c,\z\in\CC^\TFdim}\hspace{5pt} &\norm{\z}_0\; \text{ subject to }\; \syn\c\in\ASet,\,\norm{\c-\z}_2\leq\epsilon,
		\label{eq:spadeq.formulation:syn_old}\\
		\argmin_{\x\in\RR^\Tdim,\c\in\CC^\TFdim} &\norm{\c}_0\; \text{ subject to }\; \x\in\ASet,\,\norm{\x-\syn\c}_2\leq\epsilon,
		\label{eq:spadeq.formulation:syn}\\
		\argmin_{\x\in\RR^\Tdim,\c\in\CC^\TFdim} &\norm{\c}_0\; \text{ subject to }\; \x\in\ASet,\,\norm{\ana\x-\c}_2\leq\epsilon.
		\label{eq:spadeq.formulation:ana}
	\end{align}
\end{subequations}
The respective algorithms
(S-SPADQ, S-SPADQ\,DR, and A-SPADQ)
all use an iterative routine,
the principal steps of which are the adaptive hard thresholding 
and the projection onto $\ASet$.
Actually, the three SPADQ algorithms are identical to their counterparts for audio declipping,
only the definition of the feasible set $\ASet$ differs between SPADQ and SPADE.
Due to this fact and due to a~lack of space,
we do not reprint the algorithms here;
the reader can find them in \cite{ZaviskaRajmicMokryPrusa2019:SSPADE_ICASSP}.




\subsection{Inconsistent \texorpdfstring{$\ell_1$}{l1} minimization}
\label{ssec:fista.syn}	
The last option to approximate \eqref{eq:sparsity.formulation:anasyn} combines the $\ell_1$ relaxation
with the relaxation of the constraints $\x\in\ASet$ or $\c\in\SSet$.
As a~result of the constraint relaxation, the method is not consistent in the sense described above.
However, it may yield a sparser solution while
not being too far from the feasible set.
The synthesis and analysis formulations read
\begin{subequations}
	\label{eq:fista.formulation:anasyn}
	\begin{align}
		&\argmin_{\c\in\CC^\TFdim} \lambda\norm{\c}_1
		 + \frac{1}{2} \,d_{\ASet}^2(\syn\c),
		\label{eq:fista.formulation:syn}\\
		&\argmin_{\x\in\RR^\Tdim}\lambda\norm{\ana\x}_1 + \frac{1}{2}\, d_\ASet^2(\x),
		\label{eq:fista.formulation:ana}
	\end{align}
\end{subequations}
where the symbol $d_C(\cdot)$ denotes the distance from the set $C$, i.e.,
$d_C(\cdot) = \norm{\cdot - \proj_{C}(\cdot)}_2$,
and $\lambda>0$ controls the trade-off between the sparsity and the consistency of the solution.
The synthesis variant \eqref{eq:fista.formulation:syn} can be solved via FISTA,
as shown recently in \cite{RenckerBachWangPlumbley2018:Fast.iterative.shrinkage.declip.dequant-iTwist18}.
FISTA
is a~proximal gradient method
\cite{DaubechiesDefriseMol2004:ISTA,combettes2011proximal,SiedenburgKowalskiDoerfler2014:Audio.declip.social.sparsity}),
accelerated thanks to the fact that
$\frac{1}{2}d^2_{C}$ is differentiable,
with the gradient
$\nabla \frac{1}{2}d^2_{C}(\x) = \x-\proj_{C}(\x)$
\cite{RenckerBachWangPlumbley2018:Sparse.recovery.dictionary.learning}.
The resulting algorithm is Alg.\,\ref{alg:fista}.

Alternatively, we can use the DR algorithm to solve \eqref{eq:fista.formulation:syn}.
From \cite[Example 6.65]{Beck2017:First.Order.Methods}, we know that
\begin{equation}
	\prox_{\const d^2_C / 2}(\z) = 
	\frac{1}{\const+1}\left(\const\,\proj_{C}(\z)+\z\right),
	\label{eq:prox.d}
\end{equation}
%
which is a~convex combination of a~point and its projection onto
$C$.
This formula is used as the proximal operator of the second term in \eqref{eq:fista.formulation:syn}.
The resulting algorithm is Alg.\,\ref{alg:DR_dequantization.FISTA.syn}.

\begin{figure}[t]
\begin{minipage}{\columnwidth}
\removelatexerror
\begin{algorithm}[H]
	\DontPrintSemicolon
	\SetAlgoVlined
	\small
	\KwIn{Set starting points $\c^{(0)} \in \CC^\TFdim$ and $\z^{(0)}=\c^{(0)}$.\\
		Set parameters $\mu$ and $t^{(0)}=1$.
	}
	\For{$i=0,1,\dots$\,}{ 
		$\c^{(i+1)} = \soft_{\lambda\mu}\left( \z^{(i)}-\mu\ana(\syn\z^{(i)} - \proj_{\ASet}(\syn\z^{(i)}))\right)\hspace{-2em}$\\
		$t^{(i+1)} = \left(1+\sqrt{1+4(t^{(i)})^2}\right)\hspace{-0.2em}/2$\\
		$\z^{(i+1)} = \c^{(i+1)} + \frac{t^{(i)}-1}{t^{(i+1)}}\left(\c^{(i+1)}-\c^{(i)}\right)$
	} 
	\KwRet{\(\syn\c^{(i+1)}\)}
	\caption{\mbox{FISTA solving \eqref{eq:fista.formulation:syn}}}
	\label{alg:fista}	
\end{algorithm}

\begin{algorithm}[H]
	\DontPrintSemicolon
	\SetAlgoVlined
	\small
	\KwIn{Set starting point $\z^{(0)} \in \CC^\TFdim$ and
		parameter $\gamma > 0$.
	}
	\For{$i=0,1,\dots$\,}{
		$\c^{(i)} =
		   \frac{1}{\gamma+1} (\gamma\,\proj_{\SSet}(\z^{(i)}) + \z^{(i)})$ \\
		$\z^{(i+1)} = \z^{(i)} + \soft_{\gamma\lambda}(2\c^{(i)} - \z^{(i)}) - \c^{(i)}$ \\
	}
	\KwRet{$\c^{(i)}$}
	\caption{\mbox{Douglas--Rachford algorithm solving \eqref{eq:fista.formulation:syn}}}
	\label{alg:DR_dequantization.FISTA.syn}
\end{algorithm}
\end{minipage}
\vspace{-1.5em}
\end{figure}

The analysis-based problem \eqref{eq:fista.formulation:ana} can be solved using the CP algorithm;
see Alg.\,\ref{alg:CP_dequantization:fista_like}.
Note that the update of $\p$ uses the proximal operator presented in \eqref{eq:prox.d}.
%
Finally, we propose two alternatives of tackling \eqref{eq:fista.formulation:ana} by means of another approximation.
%
First, let us apply the DR algorithm to the problem. 
The proximal operator of the distance function can again be taken from \eqref{eq:prox.d}.
The proximal operator of $\const\norm{\ana \cdot}_1=\const\norm{\cdot}_1 \circ \ana$ is problematic.
If the involved linear operator were the synthesis, the same composition rule could be followed as in the case of
the projection
\eqref{eq:proj.on.sset}.
For $\ana$ being the analysis,
no such rule can be applied, though.
Nevertheless, \cite{MokryRajmic2020:Approximal.operator} shows that
an approximation can be done using the so-called \emph{approximal operator},
which turned out to be very successful in the case of audio inpainting.
In our case, it takes the form
\begin{equation}
	\approx_{\const\norm{\cdot}_1\circ\ana}(\x) = \syn \soft_{\const}(\ana\x).  	
	\label{eq:approx}
\end{equation}
Substituting the proximal operator of distance in Alg.\,\ref{alg:DR_dequantization.FISTA.syn}
with the approximal operator from \eqref{eq:approx} results in Alg.\,\ref{alg:DR_dequantization.FISTA.ana}.
As the second alternative, FISTA can also be used for the approximation of the analysis variant \eqref{eq:fista.formulation:ana},
since $d_{\ASet}$ is differentiable and the proximal operator of $\norm{\ana \cdot}_1$
can be substituted with the approx, just as above.
The resulting algorithm is in Alg.\,\ref{alg:fista.ana}.

\begin{figure}[t]
\begin{minipage}{\columnwidth}
\removelatexerror
\begin{algorithm}[H]
	\DontPrintSemicolon
	\SetAlgoVlined
	\small
	\KwIn{Set starting points $\p^{(0)} \in \RR^\Tdim, \c^{(0)} \in \CC^\TFdim$. \\
		Set parameters $\zeta, \sigma > 0$, $\zeta\sigma\norm{\ana}^2<1$, and $\rho\in[0,1]$.
	}
	\For{$i=0,1,\dots$\,}{ 
		$\c^{(i+1)} = \mathrm{clip}_{\lambda}(\c^{(i)}+\sigma \ana \x^{(i)})$\\
		$\u^{(i+1)} = \p^{(i)}-\zeta \syn{\c}^{(i+1)}$\quad  \% auxiliary\\
		$\p^{(i+1)} = \frac{1}{\zeta+1}
			\left( \zeta \, \proj_{\ASet}(\u^{(i+1)}) + \u^{(i+1)} \right)$\\
		$\x^{(i+1)} = \p^{(i+1)} + \rho (\p^{(i+1)} - \p^{(i)}) $\\
	} 
	\KwRet{\(\p^{(i+1)}\)}
	\caption{\mbox{Chambolle--Pock algorithm solving \eqref{eq:fista.formulation:ana}}}
	\label{alg:CP_dequantization:fista_like}	
\end{algorithm}

\begin{algorithm}[H]
	\DontPrintSemicolon
	\SetAlgoVlined
	\small
	\KwIn{Set starting point $\u^{(0)} \in \RR^\Tdim$ and
		parameter $\gamma > 0$.
	}
	\For{$i=0,1,\dots$\,}{
		$\x^{(i)} =
		\frac{1}{\gamma+1} (\gamma\,\proj_{\ASet}(\u^{(i)}) + \u^{(i)})$ \\
		$\u^{(i+1)} = \u^{(i)} + \syn\soft_{\gamma\lambda}\left(\ana(2\x^{(i)} - \u^{(i)})\right) - \x^{(i)}$ \\
	}
	\KwRet{$\x^{(i)}$}
	\caption{\mbox{Douglas--Rachford alg.\ approximating \eqref{eq:fista.formulation:ana}}}
	\label{alg:DR_dequantization.FISTA.ana}
\end{algorithm}

\begin{algorithm}[H]
	\DontPrintSemicolon
	\SetAlgoVlined
	\small
	\KwIn{Set starting points $\x^{(0)} \in \RR^\Tdim$ and $\u^{(0)}=\x^{(0)}$.\\
		Set parameters $\mu$ and $t^{(0)}=1$.
	}
	\For{$i=0,1,\dots$\,}{ 
		$\x^{(i+1)} = \syn\soft_{\mu\lambda}\left(\ana( \u^{(i)}-\mu(\u^{(i)}-\proj_{\ASet}(\u^{(i)}) ))\right)\hspace{-2em}$\\
		$t^{(i+1)} = \left(1+\sqrt{1+4(t^{(i)})^2}\right)\hspace{-0.2em}/2$\\
		$\u^{(i+1)} = \x^{(i+1)} + \frac{t^{(i)}-1}{t^{(i+1)}}\left(\x^{(i+1)}-\x^{(i)}\right)$
	} 
	\KwRet{\(\x^{(i+1)}\)}
	\caption{\mbox{FISTA approximating \eqref{eq:fista.formulation:ana}}}
	\label{alg:fista.ana}	
\end{algorithm}
\end{minipage}
\vspace{-1.5em}
\end{figure}

\vspace*{-.5ex}
\section{Experiments and results}
\label{sec:experiments}
\vspace*{-1ex}

For the experiments, an audio database containing ten musical excerpts sampled at 44.1\,kHz, with an approximate duration of 7~seconds, was used.
The excerpts were extracted from the EBU SQAM database\footnote{https://tech.ebu.ch/publications/sqamcd} and thoroughly selected to cover a~wide range of signal sparsity.

The signals were first peak-normalized
and then quantized according to \eqref{eq:uniform_quantization}, using 7 different
word lengths, $w = 2, 3,\dots, 8$.
The Discrete Gabor transform (DGT) was chosen as the sparsity-promoting transform, using a~Hann window 8192 samples long (185.8\,ms).
The DGT used a 75\% overlap of the windows and 16384 frequency channels.
The algorithms were implemented in MATLAB 2019b.
They rely on the \mbox{LTFAT} toolbox \cite{LTFAT} for the time-frequency
operators.

The physical similarity of waveforms was measured using the \dsdr{},
which expresses the signal-to-distortion ratio (SDR) improvement of the quantized signal to the reconstructed signal.
Note that in dequantization, the SDR is equivalent to the signal-to-artifacts ratio (SAR).
Since we are interested mostly in perceptual quality (which may not correspond to the SDR values),
we evaluate the reconstructed signals using the PEMO-Q metric \cite{Huber:2006a},
which uses an objective difference grade (ODG) scale of $-4$ to $0$ (worst to best).

\begin{figure}[t!]%
	\centering
%
%
\definecolor{mycolor1}{rgb}{0.00000,0.44700,0.74100}%
\definecolor{mycolor2}{rgb}{0.85000,0.32500,0.09800}%
\definecolor{mycolor3}{rgb}{0.92900,0.69400,0.12500}%
\definecolor{mycolor4}{rgb}{0.49400,0.18400,0.55600}%
\definecolor{mycolor5}{rgb}{0.46600,0.67400,0.18800}%
\definecolor{mycolor6}{rgb}{0.30100,0.74500,0.93300}%
\definecolor{mycolor7}{rgb}{0.63500,0.07800,0.18400}%
\begin{tikzpicture}[scale=0.58]

\begin{axis}[%
width=5.372in,
height=2.35in,
at={(0.758in,0.481in)},
scale only axis,
bar shift auto,
ybar = 1pt,
xmin=1.5,
xmax=8.5,
xtick={2, 3, 4, 5, 6, 7, 8},
xlabel style={font=\Large\color{white!15!black}},
xticklabel style={font=\large},
ymin=3,
ymax=9,
ytick={3, 4, 5, 6, 7, 8, 9},
ylabel style={font=\Large\color{white!15!black}},
ylabel={\dsdr{} (dB)},
ylabel shift = -0.1em,
yticklabel style={font=\large},
axis background/.style={fill=white},
xmajorgrids,
ymajorgrids,
]
\addplot[ybar, bar width=0.058, fill=black, draw=black, area legend] table[row sep=crcr] {%
2	0\\
3	0\\
4	0\\
5	0\\
6	0\\
7	0\\
8	0\\
};

\addplot[ybar, bar width=0.058, fill=mycolor1, draw=black, area legend, postaction={pattern=north east lines, pattern color=white}] table[row sep=crcr] {%
2	7.85103471517752\\
3	7.56430699002411\\
4	7.33419028380075\\
5	7.07217394683348\\
6	6.3787413659676\\
7	5.49101170352001\\
8	4.50266695079942\\
};

\addplot[ybar, bar width=0.058, fill=mycolor1, draw=black, area legend] table[row sep=crcr] {%
2	7.83656646525359\\
3	7.90391744159086\\
4	7.94252124135609\\
5	7.73960360511382\\
6	6.8714588641189\\
7	5.83781490324544\\
8	4.75515453402599\\
};

\addplot[ybar, bar width=0.058, fill=mycolor2, draw=black, area legend] table[row sep=crcr] {%
2	5.58529342170117\\
3	6.74855366245436\\
4	7.66292473510919\\
5	8.34843032301924\\
6	7.20469296237519\\
7	5.61956674071081\\
8	4.06775460342302\\
};

\addplot[ybar, bar width=0.058, fill=mycolor2, draw=black, area legend, postaction={pattern=north east lines, pattern color=black!65!white}] table[row sep=crcr] {%
2	6.66159488612942\\
3	7.46081115964680\\
4	8.07900481095853\\
5	8.63106586438690\\
6	7.62375027069112\\
7	6.12713901822904\\
8	4.55673054025738\\
};

\addplot[ybar, bar width=0.058, fill=black!65!white, draw=black, area legend, postaction={pattern=north east lines, pattern color=mycolor2}] table[row sep=crcr] {%
2	5.78231353458534\\
3	6.77294821148527\\
4	7.69745822873600\\
5	8.27191813694427\\
6	7.14643648349822\\
7	5.58941606254737\\
8	3.98742782495684\\
};

\addplot[ybar, bar width=0.058, fill=white, draw=black, area legend, postaction={pattern=north east lines, pattern color=mycolor3}] table[row sep=crcr] {%
2	7.39434416237511\\
3	7.03843152572199\\
4	6.91065029105854\\
5	6.70537058698311\\
6	6.09288085752758\\
7	5.36609936010957\\
8	4.45990090207856\\
};

\addplot[ybar, bar width=0.058, fill=mycolor3, draw=black, area legend, postaction={pattern=north east lines, pattern color=white}] table[row sep=crcr] {%
2	7.83794876329666\\
3	7.53973065264315\\
4	7.30500926477505\\
5	7.058707965147\\
6	6.37706305746025\\
7	5.48989226675171\\
8	4.50225169264853\\
};

\addplot[ybar, bar width=0.058, fill=mycolor5, draw=black, area legend] table[row sep=crcr] {%
2	7.82405976761063\\
3	7.87510286076394\\
4	7.89878479132963\\
5	7.73638899315282\\
6	6.8699195156623\\
7	5.83691324589429\\
8	4.7548177848856\\
};

\addplot[ybar, bar width=0.058, fill=mycolor5, draw=black, area legend, postaction={pattern=north east lines, pattern color=white}] table[row sep=crcr] {%
2	7.79926573928951\\
3	7.77764103512053\\
4	7.75744217913124\\
5	7.6840162317806\\
6	6.85297097591229\\
7	5.83388290404603\\
8	4.75479517954923\\
};

\addplot[ybar, bar width=0.058, fill=white, draw=black, area legend, postaction={pattern=north east lines, pattern color=mycolor5}] table[row sep=crcr] {%
2	7.23999048512205\\
3	7.09903553122874\\
4	7.3136628424508\\
5	7.15343384069337\\
6	6.26837689903086\\
7	5.52346069914296\\
8	4.59584565873135\\
};

\end{axis}
\end{tikzpicture}%
\\%
%
%
\definecolor{mycolor1}{rgb}{0.00000,0.44700,0.74100}%
\definecolor{mycolor2}{rgb}{0.85000,0.32500,0.09800}%
\definecolor{mycolor3}{rgb}{0.92900,0.69400,0.12500}%
\definecolor{mycolor4}{rgb}{0.49400,0.18400,0.55600}%
\definecolor{mycolor5}{rgb}{0.46600,0.67400,0.18800}%
\definecolor{mycolor6}{rgb}{0.30100,0.74500,0.93300}%
\definecolor{mycolor7}{rgb}{0.63500,0.07800,0.18400}%
\begin{tikzpicture}[scale=0.58] 

\begin{axis}[%
width=5.372in,
height=2.35in,
at={(0.758in,0.481in)},
scale only axis,
bar shift auto,
ybar = 1pt,
xmin=1.5,
xmax=8.5,
ylabel shift = -0.5em,
xtick={2, 3, 4, 5, 6, 7, 8},
xlabel style={font=\Large\color{white!15!black}},
xticklabel style={font=\large},
xlabel={word length (bps)},
ymin=-4,
ymax=-1,
ytick={-4, -3, -2, -1, 0},
ylabel style={font=\Large\color{white!15!black}},
ylabel={PEMO-Q ODG},
yticklabel style={font=\large},
ylabel shift = -1em,
axis background/.style={fill=white},
xmajorgrids,
ymajorgrids,
legend columns=2,
legend style={at={(0.02, 0.98)}, anchor=north west, legend cell align=left, align=left, draw=white!15!black, font=\large},
]
\addplot[ybar, bar width=0.058, fill=black, draw=black, area legend] table[row sep=crcr] {%
2	-3.83954692015723\\
3	-3.83524287752205\\
4	-3.81651888983447\\
5	-3.77431623468317\\
6	-3.68070064628702\\
7	-3.53164549384489\\
8	-3.3011534564243\\
};
\addlegendentry{quantized}

\addplot[ybar, bar width=0.058, fill=mycolor1, draw=black, area legend, postaction={pattern=north east lines, pattern color=white}] table[row sep=crcr] {%
2	-3.87554115680034\\
3	-3.81587171505508\\
4	-3.69567864666857\\
5	-3.52525896646116\\
6	-3.31092683935115\\
7	-2.74431594400699\\
8	-1.70478575491049\\
};
\addlegendentry{DR (Alg.\,1)}

\addplot[ybar, bar width=0.058, fill=mycolor1, draw=black, area legend] table[row sep=crcr] {%
2	-3.90030950365089\\
3	-3.83436045294789\\
4	-3.69489769744346\\
5	-3.36136798521223\\
6	-2.92979310877342\\
7	-2.24686764680157\\
8	-1.4165653537732\\
};
\addlegendentry{CP (Alg.\,2)}

\addplot[ybar, bar width=0.058, fill=mycolor2, draw=black, area legend] table[row sep=crcr] {%
2	-3.81972949614135\\
3	-3.74339733775552\\
4	-3.59055508308787\\
5	-3.34642942499600\\
6	-2.98472263692645\\
7	-2.53241278054415\\
8	-2.10469734689202\\
};
\addlegendentry{A-SPADQ}

\addplot[ybar, bar width=0.058, fill=mycolor2, draw=black, area legend, postaction={pattern=north east lines, pattern color=black!65!white}] table[row sep=crcr] {%
2	-3.77587855528543\\
3	-3.73958746807058\\
4	-3.62922045107021\\
5	-3.48861901943179\\
6	-3.14098082221167\\
7	-2.63016174622996\\
8	-2.04591513605044\\
};
\addlegendentry{S-SPADQ}

\addplot[ybar, bar width=0.058, fill=black!65!white, draw=black, area legend, postaction={pattern=north east lines, pattern color=mycolor2}] table[row sep=crcr]{%
2	-3.80434397074263\\
3	-3.73488831425455\\
4	-3.60939543282653\\
5	-3.43139983135917\\
6	-3.10577039715490\\
7	-2.66482958054587\\
8	-2.25571042536338\\
};
\addlegendentry{S-SPADQ DR}

\addplot[ybar, bar width=0.058, fill=white, draw=black, area legend, postaction={pattern=north east lines, pattern color=mycolor3}] table[row sep=crcr] {%
2	-3.90571562272665\\
3	-3.84059256056017\\
4	-3.71845273051343\\
5	-3.57062364269086\\
6	-3.35487975430941\\
7	-2.95959303100296\\
8	-2.22196904057706\\
};
\addlegendentry{FISTA (Alg.\,3)}

\addplot[ybar, bar width=0.058, fill=mycolor3, draw=black, area legend, postaction={pattern=north east lines, pattern color=white}] table[row sep=crcr] {%
2	-3.88239737792084\\
3	-3.81968343338695\\
4	-3.69328602557537\\
5	-3.52517554591176\\
6	-3.30910535924469\\
7	-2.74086908115571\\
8	-1.70319685798536\\
};
\addlegendentry{DR (Alg.\,4)}

\addplot[ybar, bar width=0.058, fill=mycolor5, draw=black, area legend] table[row sep=crcr] {%
2	-3.90787415082013\\
3	-3.84091771848031\\
4	-3.69925940714381\\
5	-3.35843730078121\\
6	-2.92892130955126\\
7	-2.23108929806046\\
8	-1.41536818715546\\
};
\addlegendentry{CP (Alg.\,5)}

\addplot[ybar, bar width=0.058, fill=mycolor5, draw=black, area legend, postaction={pattern=north east lines, pattern color=white}] table[row sep=crcr] {%
2	-3.91642032450482\\
3	-3.85061257881407\\
4	-3.71403247082553\\
5	-3.36490427426282\\
6	-2.93704382852469\\
7	-2.23304820915532\\
8	-1.41551157695146\\
};
\addlegendentry{DR (Alg.\,6)}

\addplot[ybar, bar width=0.058, fill=white, draw=black, area legend, postaction={pattern=north east lines, pattern color=mycolor5}] table[row sep=crcr] {%
2	-3.92811927070694\\
3	-3.85195095056072\\
4	-3.71851265108876\\
5	-3.48197969679086\\
6	-2.98277095176228\\
7	-2.33197807032947\\
8	-1.66497097252712\\
};
\addlegendentry{FISTA (Alg.\,7)}

\end{axis}
\end{tikzpicture}%
%
	\vspace{-1em}
	\caption{Average \dsdr{} and PEMO-Q ODG results.}%
	\vspace{-1em}
	\label{fig:dSDR_pemoq_bars}%
\end{figure}
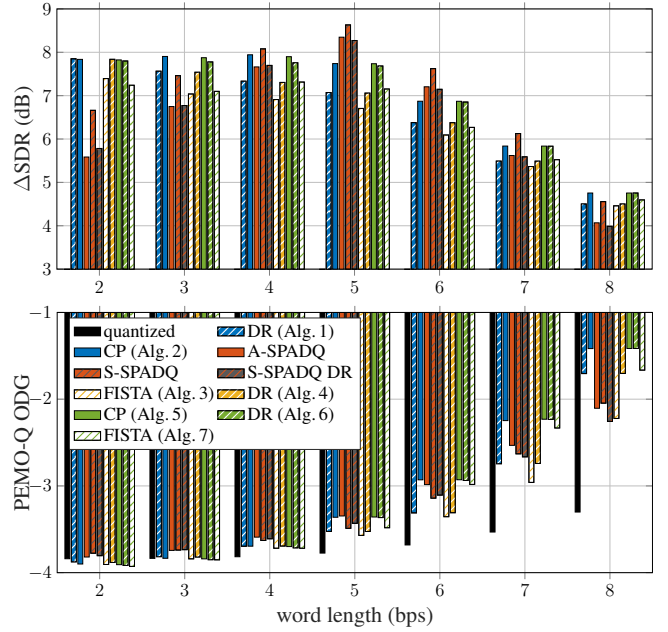

The parameters of the algorithms need fine-tuning to achieve the best possible results.
In the case of
$\ell_1$ minimization, the \dsdr{} values tend to gain rapidly during the first couple of iterations but then drop and stabilize at a~lower value.
This is explained by the fact that during the convergence,
the algorithms retain the signal within the consistent area,
while the $\ell_1$ norm of the coefficients decreases.
When their $\ell_1$ norm is pushed too far towards zero, the waveform is also affected, tending to incorrectly settle close to the edge of the feasible quantization intervals.
Interrupting the convergence at the SDR peak provides results with the most similar waveforms to the original
(in practice unknown) signal.
The first bar chart in Fig.\,\ref{fig:dSDR_pemoq_bars} shows the best achievable \dsdr{} values,
which in our case correspond to stopping the convergence after approximately 100 iterations.
Letting the algorithms fully converge yields significantly better results in terms of the perceptual metric.
The PEMO-Q ODG values are presented as the second chart of Fig.\,\ref{fig:dSDR_pemoq_bars} and they were reached after 500 iterations of each algorithm.

The \dsdr{} results suggest no clear winner.
The SPADQ algorithms perform well for word lengths of 4--7 bps but for other tested word lengths they are outperformed by the convex methods.
In the case of consistent $\ell_1$ minimization, it is clear that except for the 2~bps case,
the analysis variant using the CP algorithm outperforms the synthesis variant using the DR algorithm. 
The results of the inconsistent problem formulations also indicate the predominance of analysis-based formulations; this behavior can also be observed in audio declipping and inpainting. 
The effect of inexact computation of the thresholding step \eqref{eq:approx} turns out to have negligible influence,
as in the case of inpainting \cite{MokryRajmic2020:Approximal.operator}.

The PEMO-Q results indicate that for $w \geq 4$~bps all the methods improve the perceptual quality. 
The overall \mbox{PEMO-Q} results roughly correspond to the \dsdr{} results. 
Small differences can be found in the case of SPADQ 
or methods based on inconsistent $\ell_1$ minimization.
Finally, the FISTA algorithms seem to perform worse than the other methods in both the \dsdr{} and PEMO-Q.

The MATLAB implementation and data are available at 
\href{http://github.com/zawi01/audio_dequantization}{http://github.com/zawi01/audio\_dequantization}.

\vspace*{-1ex}
\section{Conclusion}
\label{sec:conclusion}
\vspace*{-1ex}
The paper discussed a number of sparsity-based approaches to dequantization.
The audio signals were subjected to uniform quantization.
The signals were reconstructed
using convex and non-convex approaches,
respecting strictly, or only approximately the solution consistency. 
None of the $\dsdr{}$ and PEMO-Q results suggest a~clearly preferred method for audio dequantization.
Among the convex methods, the variants involving the analysis time-frequency operator 
appear to give better results than the synthesis-based variants do.

\clearpage


\newcommand{\noopsort}[1]{} \newcommand{\printfirst}[2]{#1}
  \newcommand{\singleletter}[1]{#1} \newcommand{\switchargs}[2]{#2#1}

\end{document}